# ESOSEG

**Environment for Simulation, Operation and Optimization of Smart Energy Grids**

# Deliverable TP2.AP2.0

# Modeling Adaptive Self-healing Systems

## Version: 1.0

| Projekt | ESOSEG | |
|---|---|---|
| Verantwortlich | TUM | |
| Review durch | Dominik Ascher | |
| Autor(en) | Habtom Kahsay Gidey, Diego Marmsoler, and Dominik Ascher | |
| Erstellt am | 01.10.2018 | |
| Letzte Änderung | 15.03.2019 | |
| Freigabe | | Konsortium (HSU, SEKAS, TUM) |
| | | Offen für Projektpartner |
| | | Öffentlich |
| Arbeitszustand | | In Bearbeitung |
| | | Bereitgestellt |
| | | Release |

# Änderungsverzeichnis

| Änderung | | | Geän-derte Kapitel | Beschreibung | Autor | Zustand |
|---|---|---|---|---|---|---|
| Nr. | Datum | Version | | | | |
| 1 | 01.10.2018 | 0.1 | alle | Ersterstellung, outline | Gidey and Marmsoler | |
| 2 | 21.11.2018 | 0.2 | | Sections 1,2 and 3 added | Gidey and Marmsoler | |
| 3 | 16.02.2019 | 0.6 | | Draft Report Review | Gidey and Ascher | |
| 4 | 28.02.2019 | 0.8 | | Changed to ESOSEG document format | Gidey and Ascher | |
| 5 | 03.03.2019 | 0.9 | | Abstract added | Gidey | |
| 6 | 12.03.2019 | 1.0 | | Version 1.0 Finalized | Gidey | |

# Abstract

**Motivation:** Smart grids design requires energy distribution operations to be adaptable to abnormality. This requirement entails distribution system operators (DSOs) to optimize restoration to normal operational states dynamically. However, these design challenges demand collaborative research efforts on sophisticated modeling and simulation approaches.

**Approach:** In the ESOSEG research project, analyzing the smart grid domain as a software-intensive system, we employed a dynamic architecture approach, particularly the FOCUS theory, to model and assure the domains' self-healing requirements. Although some works specify various self-healing systems, to the best of our knowledge, the use of the approach in smart grids is the first work to enable a formal specification and verification of self-healing properties in smart grids.

**Results:** As a result, to support the modeling and verification process, we developed tool support with Eclipse Modeling Framework (EMF), Xtext, and other languages in the EMF ecosystem. The tool includes a grammar or a meta-model of the DSL, an interface to enable textual and graphical modeling of architectural patterns and code transformer engine for verification. Furthermore, we evaluated the modeling and verification features of the tool support with an e-Car charging scenario for modeling adaptive self-healing properties.

**Futureworks:** As an outlook, future works could include investigation of comprehensive case studies. These, for instance, could be further particular adaptability scenarios addressing challenges in DSOs. Another interesting aspect could be the evaluation of the modeling approach by investigating its use with engineers involved in a smart grid design. Next, the evaluation could be followed with abstractions of the verification process to make it useable by system architects with no knowledge of the proof language, Isabelle/HOL.

# Contents

# 1  Introduction

Today, society has very high expectations for electricity distribution systems. Reliability and automatic restoration in situations of overload, under load, and all forms of unexpected failure are some of the critical electric grid design requirements (Ascher, 2017). These design challenges demand collaborative research efforts and complex design approaches. The ESOSEG (Environment for Simulation, Operation, and Optimization of Smart Energy Grids) research project is one of the organized research efforts towards actualizing advanced Smart Energy Grids (Ascher, 2017). The project is funded by the German Federal Ministry for Economic Affairs and Energy. It aims at building an open platform for the communication of information systems of distribution system operators.

Attempts to realize modern electric grids addressing these challenges are named Smart Grids or Smart Energy Grids. Smart Grids can be described to endow properties, as a result of increased digitization, dynamic optimization, and adaptive self-healing capabilities. A self-healing system capability is the property of a system to repair, restore and recover its disturbed or disrupted behavior to a specific or required mode of operation autonomously (Rodosek, 2009). Further, the provision of self-healing property depends on other attributes such as self-adaptation, which include autonomous self-monitoring and reconfiguration. Other than that, a system's context or domain is also highly determinant on the particular aspects of the dynamic decisions as adaptive self-healing capabilities.

In the ESOSEG project, after analyzing participating partners' context or domain, we related the problems of modeling the dynamic properties of adaptive self-healing models as challenges that could be addressed by dynamic architectures. In software-intensive systems, dynamic architecture specification models and languages attempt to formally describe the structure, behavior, and evolution/change of a system at runtime. Some have already tried to apply dynamic architecture specification methods in various self-healing systems (Ghosh, 2007). In our case, we employed dynamic modeling methods based on the popular FOCUS methodology (Broy, 2014). A description of the modeling approach and tool support is provided in section 2.

To follow up, we conducted the domain and impact analysis at the initial phase of the project (Ascher, 2018; Ascher, 2017). It is a systematic study aimed at understanding the challenges in smart grid, particularly distribution system operators (DSOs). Further, we supported the analysis by an investigation of the state of practice conducting interviews and requirement analysis workshops. It was also complemented by a literature review to investigate the state-of-the-art research in modeling adaptive self-healing systems. The process of analysis is illustrated in Fig. 1.1.

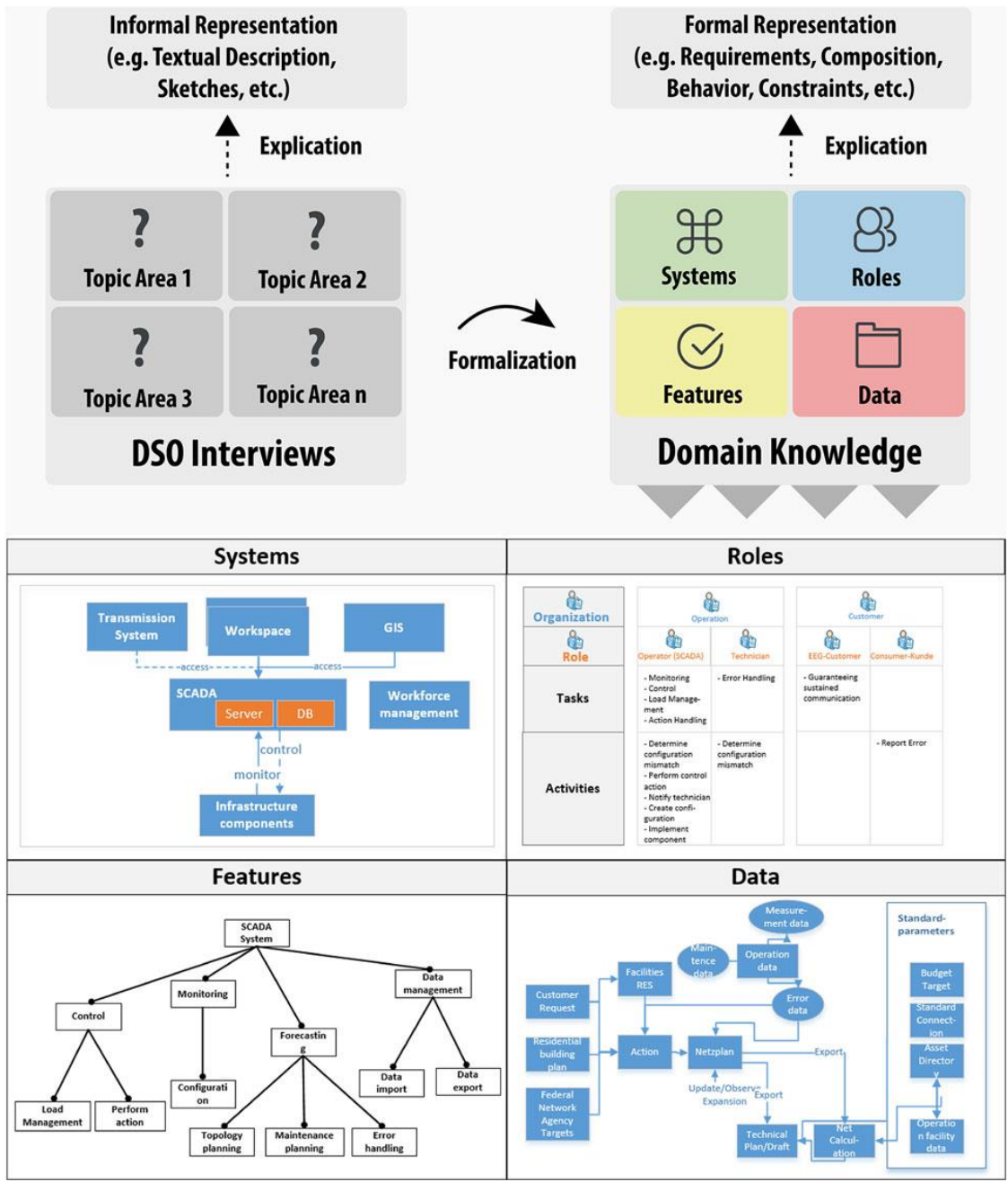

*Figure 1.1: Systematic Study of the Domain Model Analysis (Ascher, 2017)*

As a result, in this document, we report our deliverables, an approach and a tool support framework that could be utilized to model modeling adaptive self-healing systems. First, Section 2 covers a comprehensive description of the tool support for the modeling approach. Next, Section 3 provides a short example of a scenario for electric car charging station switches.

# 2  Tool Support for Modeling Adaptive Self-healing Systems

Based on the methodology introduced in Section 1, we developed a Domain Specific Language (DSL) and tool support, named FACTum Studio. FACTum Studio enables specification and verification of a systems' dynamic behavior at a design or architecture level. Thus, it can be used to facilitate the development of reference architectures or schemas in software-intensive systems. Largely, FACTum Studio can be used as well to model adaptation and self-healing properties of systems using architectural patterns. This statement, mainly, comes from the underlying assumptions of models in FACTum Studio: which are generic sets of a category of design propositions or patterns focusing on a distinct concern composed as building blocks of a complex hierarchy of architectural models. Here, the building blocks are Architectural Design Patterns (ADPs). ADPs aid to model a system's structure and behavior as hierarchical compositions of design concerns.

Moreover, Architectural Design Patterns (ADPs) are essential tools in software engineering to capture architectural design experience and are regarded as the 'Grand Tool' for designing software and systems architecture (Taylor, 2010). Each ADP contains essential architectural elements, such as the design constraints and design assurances it addresses. These architectural elements are the types of components inside, their interactions in terms of interfaces and ports, and their behavior in terms of configuration traces.

FACTum Studio is not only a specification tool, but it also sets the framework to support verification of the architectural and design assurances using interactive theorem proving with Isabelle/HOL. It is more of a framework that facilitates specification and verification of systems' dynamic properties, such as adaptability, for any patterns of domain models in software-intensive systems.

The tool, an Eclipse/EMF-based application, models ADPs with a specification of abstract data types, a textual and graphical representation of component types with their interfaces and ports, a textual specification of architectural constraints and guarantees. Additionally, it further provides automatic generation of Isabelle/HOL theory from the specification of a pattern model. It is developed using the Eclipse Modeling Framework (EMF), particularly, with Obeo's free and ready-to-use Eclipse package Obeo Designer

Community. It includes the technologies required for the development, such as Sirius, Xtext, and Xtend. FACTum Studio is licensed with an open source license and is publicly available on a GitHub repository.

In the following subsections, we briefly discuss the tool's architecture (2.1), how to get started with the tool (2.2) and provide an overview on how to use it with e-Car charging example self-healing property pattern model (2.3).

## 2.1 FACTum Studio - Architecture

FACTum Studio is implemented based on the Eclipse Modeling Framework (EMF), particularly the Obeo Designer Community edition (TypeFox, 2017). The development of the tool utilizes frameworks and languages in the Eclipse/EMF ecosystem, which include Xtext, Ecore, Xtend and Sirius (Bettini, 2016; TypeFox, 2017). Xtext is a language engineering framework that enables the textual feature of DSL using the base grammar within the Eclipse Modeling Framework (Bettini, 2016). On the other hand, Xtend is a programming language behind that enables the programing of additional language features such as type checking, validation, referencing and scoping. The Xtend language is an optimized form of Java for productivity and expressiveness (Bettini, 2016).

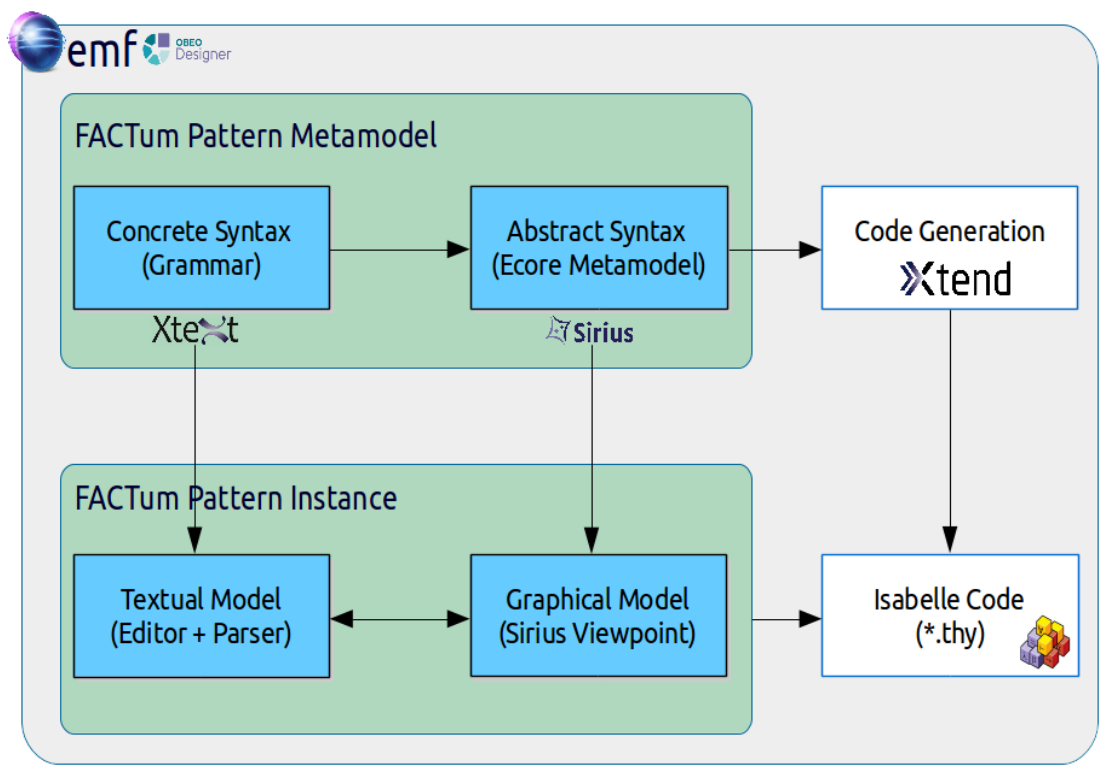

*Figure 2.1: FACTum Studio Architecture.*

Figure 2.1 illustrates the architecture of FACTum Studio. The tool has two workspaces. The first part is where the domain language is modeled and defined, which is the FACTum pattern metamodel. It is the development workspace where the grammar, constraints, validations, and the code generation templates are developed. Additionally, the workspace contains standard EMF editors that support the development of the domain metamodel. The second part, the FACTum pattern instance, is the user workspace or the runtime of the domain language, where patterns are modeled based on the metamodel. The user workspace of the language provides textual and graphical editors for the specification of pattern models. It also enables code generation from the specified model.

### 2.1.1 The Language Grammar

The grammar is the FACTum language definition, which contains all the production rules describing the concrete syntax. It includes the concrete syntax of abstract datatypes, component types (including their interfaces and behavior), architectural constraints, and architectural guarantees. All the other FACTum language infrastructure elements are mainly generated from the FACTum grammar. It includes the Ecore model, the parser and text editor, and other additional common language infrastructure code. Figure 2.2 shows an example of the production rules for the domain language elements. It sets the production rules that define the pattern language elements.

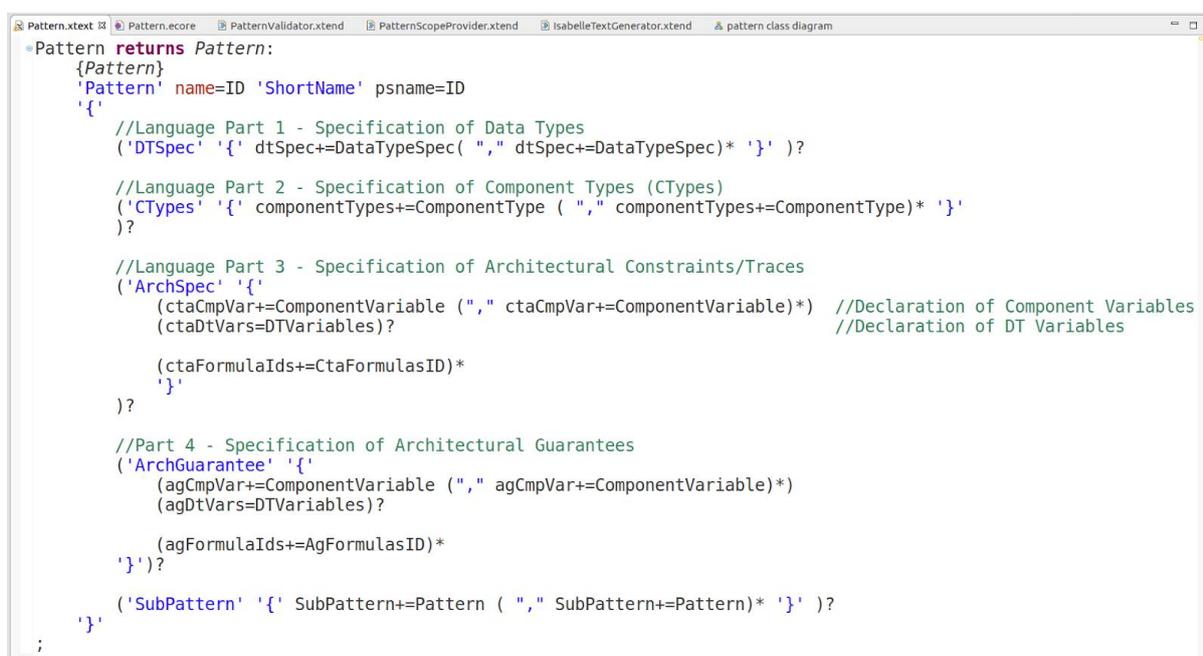

*Figure 2.2: An Example of the Language Grammar Production Rules.*

The FACTum grammar implementation is developed based on Xtext. It provides an infrastructure for parsing, linking, type checking, and an editor which works smoothly with the Eclipse-based IDE. The FACTum grammar file, Pattern.xtext, is available in the project's source code repository in GitHub (Gidey, 2018).

## 2.1.2 The Ecore Metamodel

The Ecore Metamodel is the central or core part of EMF-based languages. It describes domain models in the form of classes, attributes, and their relationships (TypeFox, 2017). Additionally, it provides a structured data model to support the generation of abstract syntax trees and a set of Java classes based on the grammar.

Figure 2.3 shows the Ecore Metamodel of FACTum Studio. It shows each elements the ecore metamodel that defines the abstract syntax tree for the domain language.

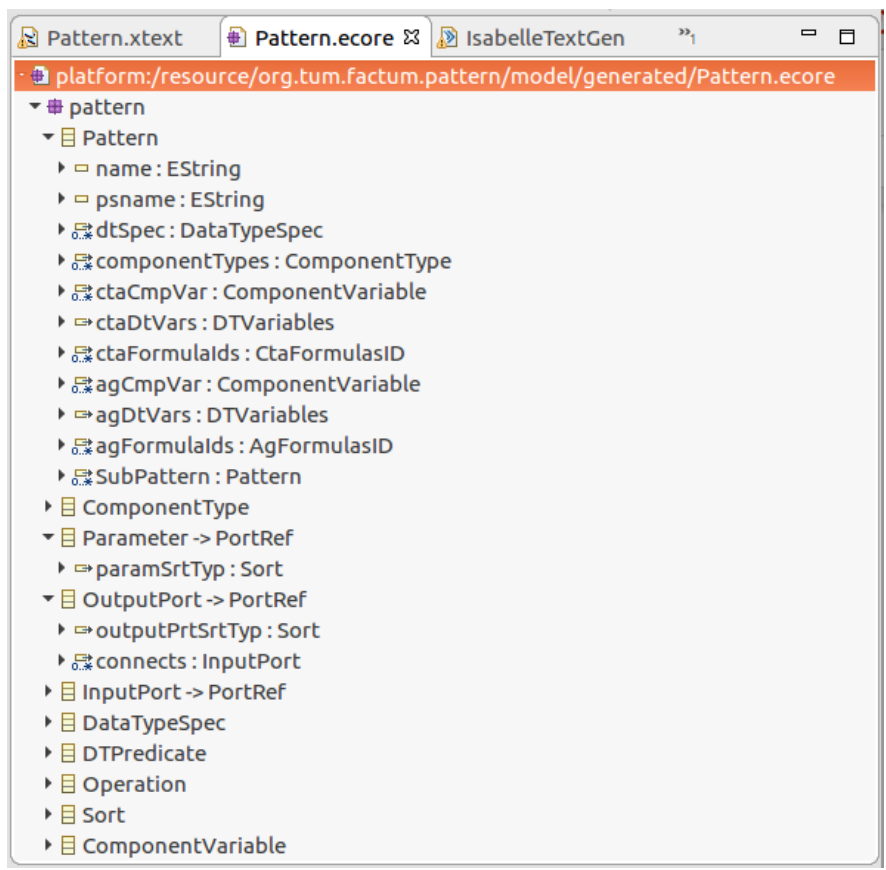

*Figure 2.3: Partial View of the Ecore Metamodel.*

In FACTum Studio, the Ecore metamodel is generated automatically from the grammar as an Xtext language artifact and named (Pattern.ecore) (Gidey, 2018). Additional language elements such as referencing, validation, constraint checks, and type checking are implemented based on the Ecore model. The implementation of constraint checks, validation, and other language support is developed using Xtend.

### 2.1.3 Code Generation

Based on the parsed Ecore model, models can be interpreted and transformed into other language codes or models. Code generation in EMF is implemented with the Xtend template engine. After the desired transformation mapping is developed, EMF automatically integrates the code generator into the runtime workspace.

Similarly, the FACTum code generator template implementation is developed, based on Xtend. It generates Isabelle/HOL theories, based on a specified pattern. The FACTum code generator template files are available in the project source code repository in GitHub (Gidey, 2018).

### 2.1.4 The Textual Model

The FACTum language textual editor provides the support to specify models of ADPs in a textual format. Starting with a definition of data types, it enables specification of all other elements of patterns, such as component types, architectural constraints, and architectural guarantees. The text editor is part of Xtext generated language artifacts. The FACTum Textual model example files are available in the project code repository (Gidey, 2018).

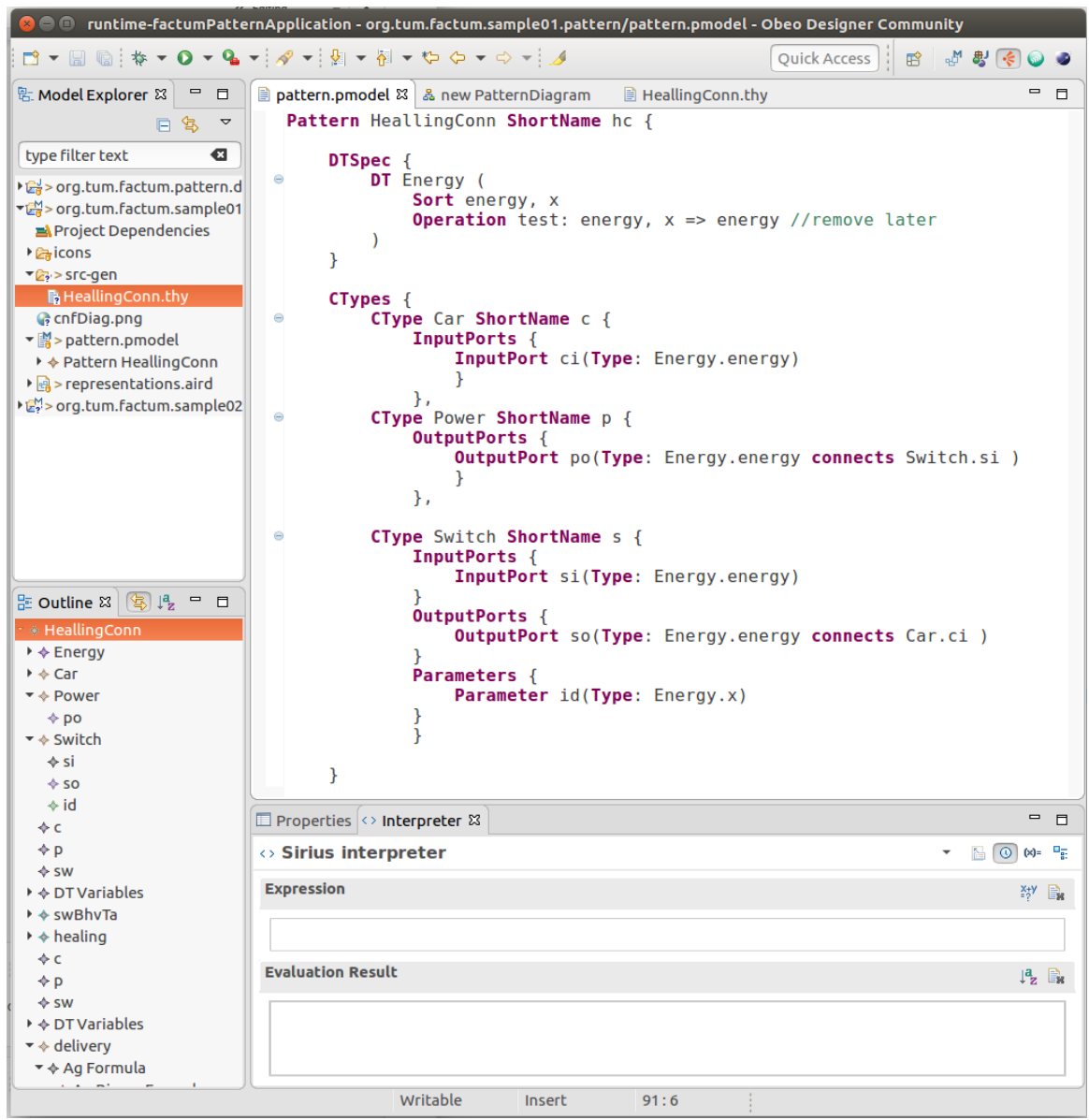

Figure 2.4: UI for the Textual Specification Language.

### 2.1.5 The Graphical Model

Similar to the text editor, the FACTum graphical editor enables the specification of pattern elements but graphically. It provides a list of graphical items in a palette, where a user can drag and create pattern elements. Currently, the graphical editor does not enable specification of behavior and constraints, which must be defined in the textual editor. The graphical editor is developed based on Sirius. It is a graphical modeling and visualization tool based on EMF. Sirius graphical editors are developed by configuring viewpoint specifications on Sirius.

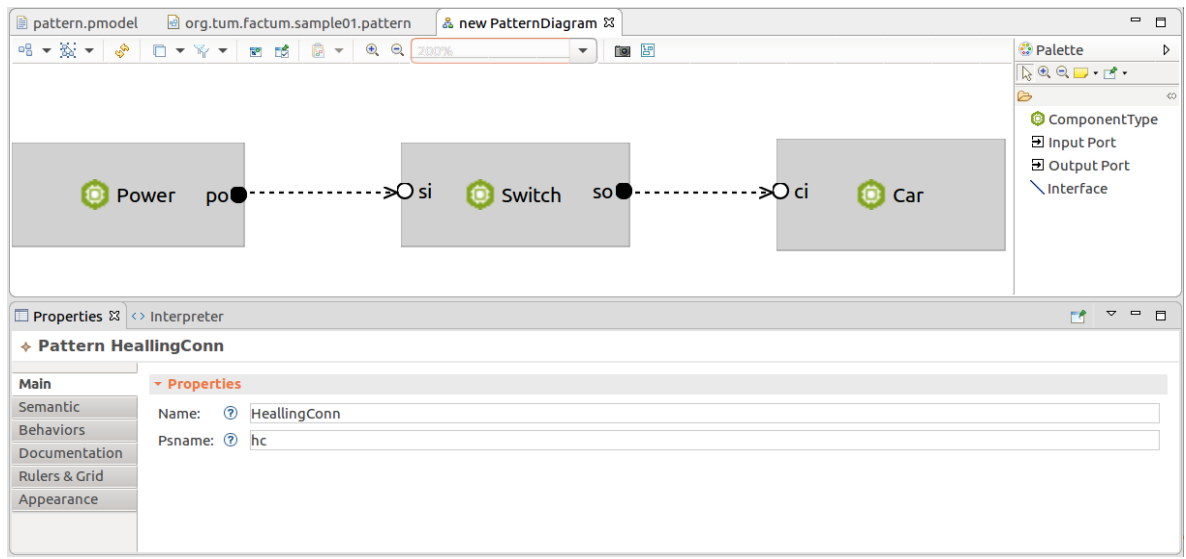

*Figure 2.5: The Graphical UI.*

Changes to the graphical model are automatically reflected in the textual model and vice versa. The synchronization is triggered, once the model is saved. The FACTum Graphical model example files are available in the project code repository (Gidey, 2018).

## 2.2 Getting Started with FACTum Studio

In the following we discuss about getting started with FACTum Studio. We will first begin by pointing out what are the required prerequisites, second, we narrate the installation instructions and then move on to how to get FACTum Studio running on a system and finally how to get it set up to start working with.

At the end, this section should give a necessary overview and resources to enable and introduce a potential user of the tool on how to set up FACTum Studio and use it.

### 2.2.1 Prerequisites:

To setup and start using FACTum Studio, a user will be required to acquire a basic understanding of the Eclipse IDE and in particular the Eclipse EMF (Bettini, 2016).

Further, a working copy of Obeo Designer Community edition [cite] is required installed to enable the importing of the FACTum tool metamodel project and runtime sample to setup and use the tool. Currently, FACTum Studio works with Obeo Designer Community Version 10.1. Other versions may not run as expected.

After the installation of Obeo Designer plugins such as Xtext and related plugins to the integration of Sirius are required to be installed and configured.

Finally, for the generated code could to be interactively verified a knowledge of Isabelle/HOL is required, however there is no need to install and understand Isabelle/HOL theory to just specify a pattern model and generate theory for verification.

### 2.2.2  Installing and Configuring FACTum Studio

To work with FACTum Studio, the Obeo Designer, FACTum Metamodel Project and Runtime Application should be downloaded, installed and configure. For demonstration purposes the FACTum Studio project files come with a sample Publisher-subscribe pattern model or specification. A video tutorial is also provided to show the specification and verification workflow with Publish-subscribe Pattern Tutorial (Gidey, 2018). The video tutorial shows an overview of using FACTum Studio with the example. It demonstrates the tool with a specification of the publish-subscribe pattern and transforms the specification to Isabelle/HOL for verification.

The following instructions guide the proper installation and configuration of FACTum Studio:

[Download Eclipse, Obeo Designer:][1]

- First, extract the downloaded Obeo Designer Community zip file into a directory where you would intend to run it.
- Next, go to the extracted directory and run the file 'obeodesigner' which is the program launcher.
- Then, when the program is open and running, install required plugins - Help menu -> Install New Software, these are Obeo Designer Community Edition Extensions or plugins such as the 'Sirius Integration with Xtext' and 'Xtext Complete SDK' from the Obeo Designer Community Edition Extensions.

[Download FACTum Metamodel Project and Runtime Application:][2]

- First, import the project file 'metamodelFACTumS.zip' into your Obeo Designer Community workspace, the project contains two archived files 'metamodelFACTumS.zip' and 'runtimeFACTumS.zip'.
- Then, generate Xtext Artifacts from the file Pattern.xtext.
- Next, configure an Eclipse runtime application (e.g., runtimeFACTumS) to launch the tool's application demo and then

---

[1] *https://www.obeodesigner.com/en/download*
[2] *https://goo.gl/fgZN2Y*

- Import the file 'runtimeFACTumS.zip' into your created Eclipse application (runtimeFACTumS) to try and test the FACTum demo.
- Verify the example code generated by copying the proof from the file 'factum/examples/PublisherSubscriber.thy' into the generated Isabelle code.

## 2.3 Key Features of FACTum Studio

At this stage, the tool's main features range over the modeling of a domain model schema architecture as an architectural pattern to transforming these models into Isabelle/HOL code for verification (Gidey, 2018).

### 2.3.1 Textual Specification of Abstract Data Types

As discussed in Section 1, data types are specified using algebraic specification techniques (Broy, 2000). At the moment abstract data types are specified only at the textual models but can be used at both the graphical and textual models. Each data type specification, may contain sorts, operations, and predicates. First, list of sorts is specified and then, the sorts can be used to specify characteristic functions or operations for a data type. Figure 2.6, shows an example Publisher-Subscriber pattern data type specification with two data type specifications, Event and Subscription. The specification Event contains the events for which subscribers may subscribe as well as corresponding messages. Event also has an operation "evt" to associate messages with events, and a predicate "in" to check whether an event is actually contained within a set of events. The specification Subscription, on the other hand, contains identifiers SID for subscriber components and the actual subscriptions SBS. Subscription also declares corresponding functions, sub and unsub, to subscribe to, and unsubscribe from events.

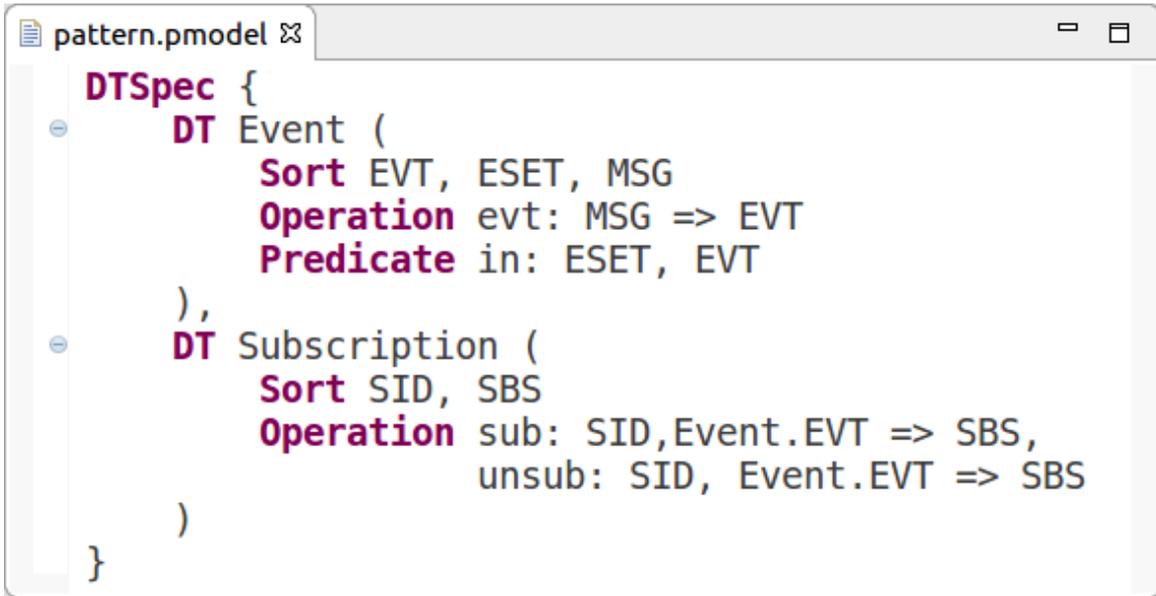

*Figure 2.6: Publisher-Subscriber Pattern Data Type Specification.*

Moreover, FACTum Studio supports the data type specification process by doing static checks to ensure sorts used in a function or predicate specification are already specified earlier. Figure 2.7, shows an example of a static check in the data type specification. It checks the existence of a sort before it can be used to specify an operation. In such cases, if a sort is not specified or a wrong sort is referenced, the tool notifies the problem and suggests a list of existing sorts to fix and address it.

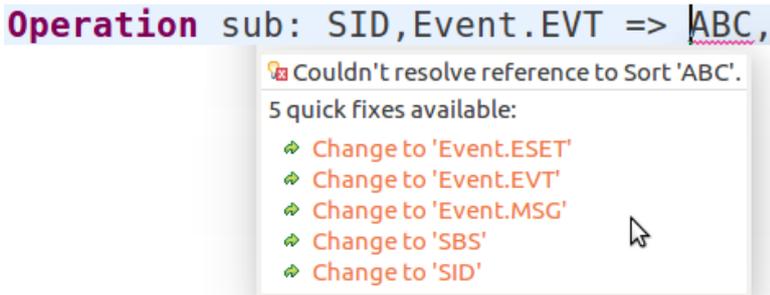

*Figure 2.7: Data Type Specification Checks and Constraints*

### 2.3.2 Graphical Specification of Architectural Diagrams

The graphical UI enables the modeling and specification of architecture diagrams elements, such as component types, input output ports and the interfacing connectors (Marmsoler, 2018, April). Component types are represented as a labeled rectangle, input ports with outlined circles, output ports solid-filled and connectors with solid or dotted lines. Figure 2.8, shows an example of an architecture diagram specification for Publisher-Subscriber pattern model. The diagram shows the Publisher and Subscriber component types, each containing an input and output port.

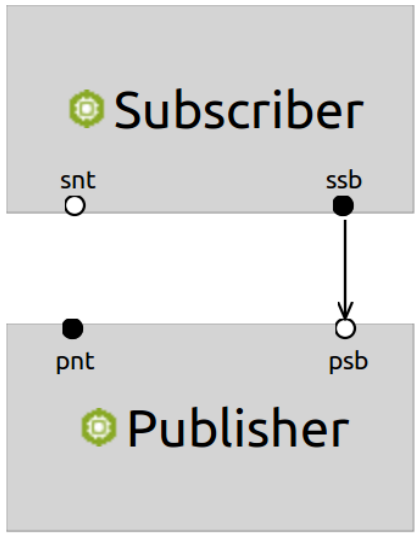

*Figure 2.8: Architecture Diagram for Publisher-Subscriber pattern*

Additionally, port data types can be specified at the graphical UI/model by setting the port properties specification which is also available to each diagram elements with varying settings. Moreover, Figure 2.9, shows an example of the property tab for the publisher's input port and demonstrates the setting of already specified data type sort to the input port. The output port and the input port of both component type are typed by already specified sorts.

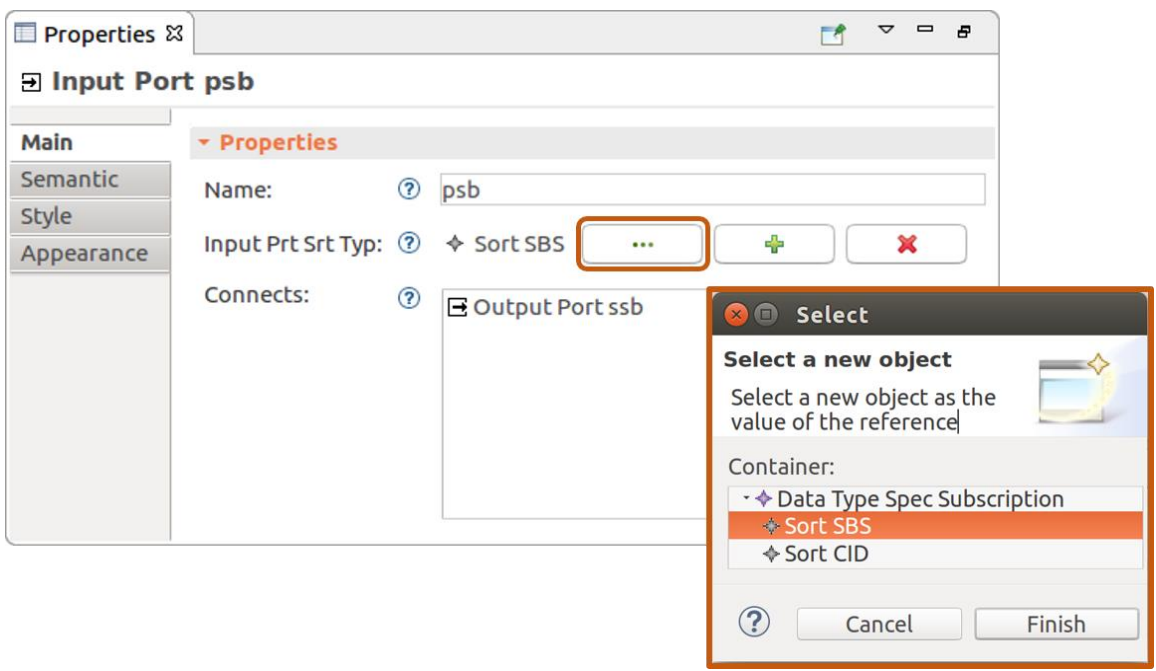

*Figure 2.9: Property Panel for Diagram Elements*

### 2.3.3  Specification of Architectural Constraints and Guarantees

FACTum Studio models constraints and guarantees using the textual specification language. Both architectural constraints and guarantees are specified using behavior and architecture trace assertions (Marmsoler, 2017). In both cases, component behaviors are specified using behavior trace assertions, which are linear temporal logic formulae with terms formulated by port names and data type functions. On the other hand, the activation and connections of components are specified using architecture trace assertions, which are similarly linear temporal logic formulae with special predicates to denote component activation and port connection. Compared to behavior trace assertions (which are specified over a single interface), architecture trace assertions are specified over an architecture (a set of components). Thus, their specification relies on the concept of component variables: variables which are interpreted by components of a certain type.

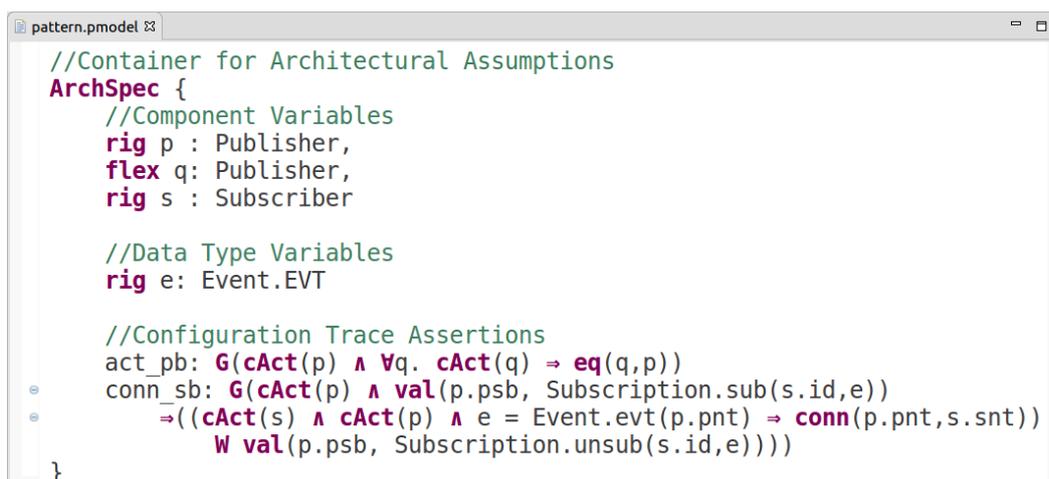

*Figure 2.10: Architectural Constraints for Publisher-Subscriber*

Figure 2.10, shows an example Publisher-Subscriber pattern architectural constraints depicting the specification of two architectural constraints. The constraint act_pb requires that a publisher component is always activated and unique and the second constraint conn_sb requires that a subscriber component is connected to the unique publisher component, whenever the latter sends out a message for which the subscriber is subscribed.

To specify the constraints, first three component variables and one datatype variable is declared. Variables could be rigid or flexible. A rigid variable keeps its value over time, and a flexible variable is newly interpreted at each point in time. Next, two constraints are specified in terms of two linear temporal logic formulae over the component and datatype variables. Moreover, the constraints formulae make use of four architecture predicates to denote valuation of component ports with certain messages (val), component activation (cAct), connection between component ports (conn), and equality of components (eq).

Similar to the specification of constraints described, architectural guarantees are specified architecture trace assertions. Figure 2.11, shows an example Publisher-Subscriber pattern architectural guarantees depicting an architectural guarantee specification, named delivery. The guarantee ensures that a subscriber indeed receives all the messages for which it is subscribed.

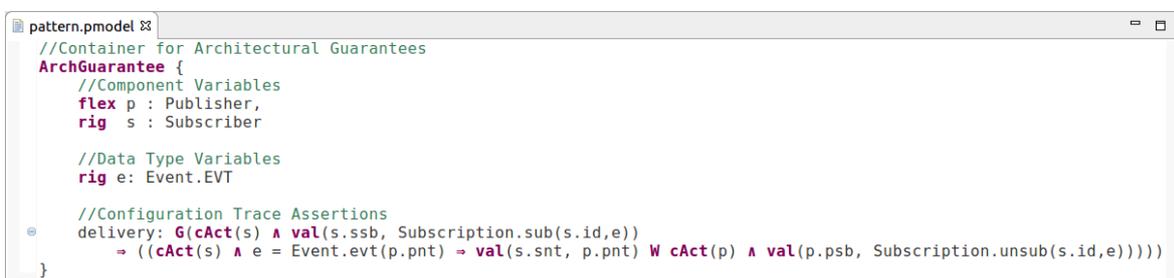

Figure 2.11: Architectural Guarantee for Publisher-Subscriber

Moreover, Figure 2.12, shows an example of a feature which conveniently checks for the specification language constraints and signature checks with the constraints and guarantees specification. One convenient feature, which is demonstrated at the top of the figure, is its support to specify component ports: whenever such a port is used in the specification, the tool ensures that the port is indeed consistent with the type of the corresponding component variable. Thus, since variable "p" is of type Subscriber, only ports "pnt" and "psb", declared for a subscriber interface, can be referenced.

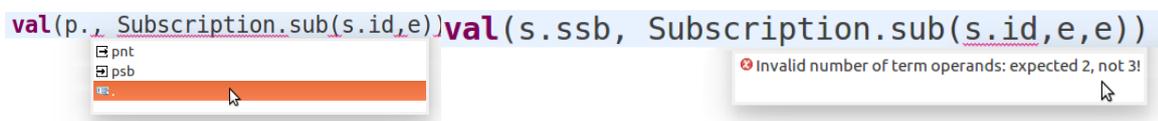

Figure 2.12: Specification Language Constraints and Signature Checks

Another convenient feature is the signature check, as demonstrated at the second part of the figure: whenever we use a datatype function, FACTum Studio ensures that the parameters passed to the function (either port values or other functions) are indeed consistent with the function's signature as specified in the corresponding datatype specification. Thus, in the example, since function "sub" was declared to take two parameters only, we are indeed not allowed to pass more than two parameters to "sub".

### 2.3.4  Generation of Isabelle/HOL Theories

Whenever specified model is valid and saved FACTum Studio generates Isabelle/HOL theory code automatically. The theory consists of an Isabelle/HOL locale specification (Ballarin, 2003), which contains corresponding assumptions for each architectural constraint. The pattern's guarantee is used to generate a corresponding Isabelle/HOL theorem for the locale which can then be proved using a calculus for dynamic architectures (Marmsoler, 2017, October). Figure 2.13, shows an example of generated Isabelle/HOL theory. The mapping and corresponding proof is provided online in the FACTum Studio GitHub repository.

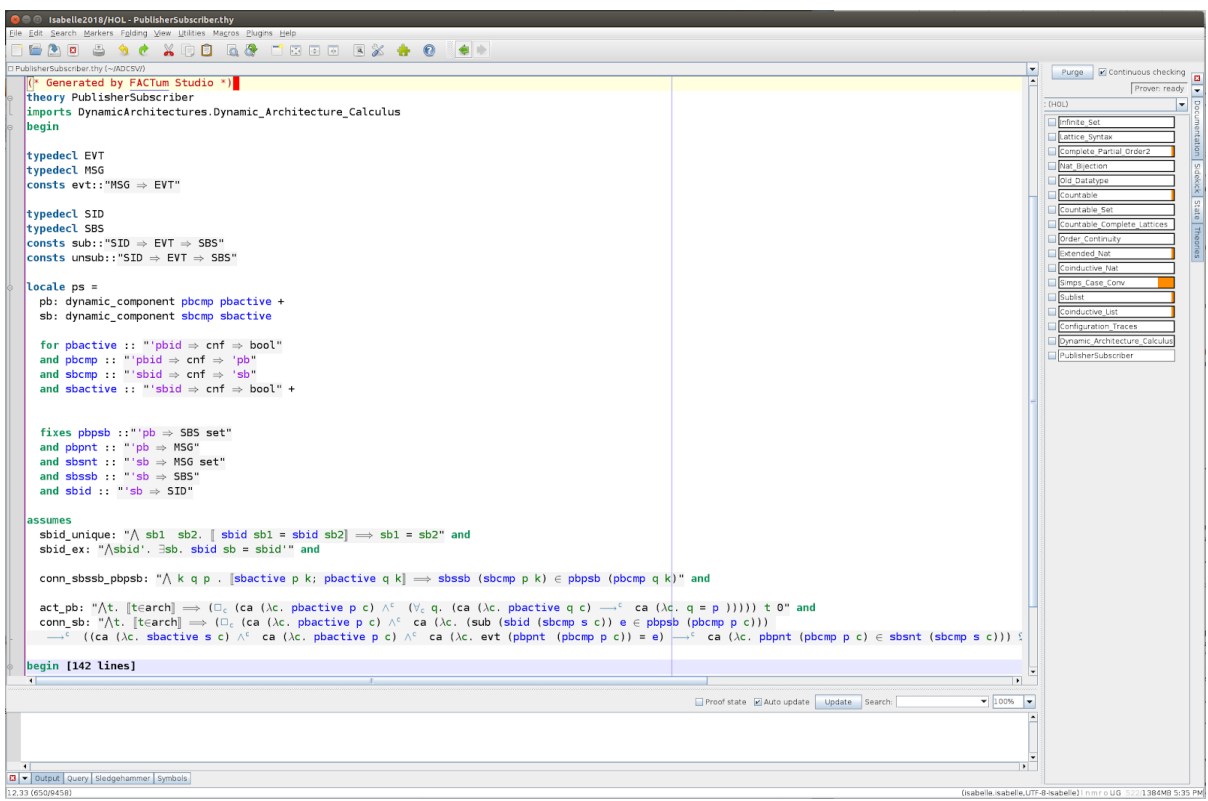

*Figure 2.13: An Example of Generated Isabelle/HOL Theory*

## 3  Self-healing Scenario - Electric Car Charging

Figure 3.1 illustrates an example an adaptive self-healing model of a charging station switch. It shows self-healing property pattern model. The example is simplified on purpose to demonstrate the modeling process with a small scenario possible. The model specifies three component types, Power station representing charging station points, a switch, and a charging e-car. The Power component type has an output port connected to a switch. The Switch component has an input port and an output port connected to an input port of the Car component type.

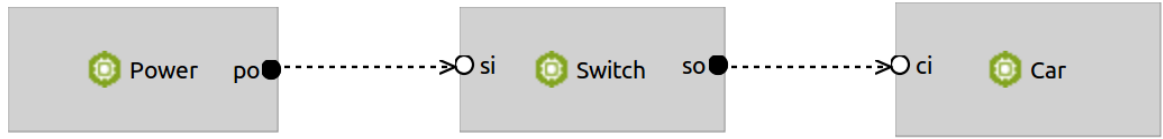

*Figure 3.1: Self-healing Switch Architecture Diagram.*

```
Pattern HeallingConn ShortName hc {

    DTSpec {
        DT Energy (
            Sort energy
        )
    }

    CTypes {
        CType Car ShortName c {
            InputPorts { InputPort ci(Type: Energy.energy) }
        },
        CType Power ShortName p {
            OutputPorts { OutputPort po(Type: Energy.energy connects Switch.si) }
        },
        CType Switch ShortName s {
            InputPorts { InputPort si(Type: Energy.energy) }
            OutputPorts { OutputPort so(Type: Energy.energy connects Car.ci) }
        }
    }

    //Container for Architectural Constraints
    ArchSpec {
        flex c : Car,
        flex p : Power,
        flex sw : Switch

        //Data Type Variables
        rig e: Energy.energy

        //Configuration Trace Assertions
        swBhvTa: (G(val(sw.si, e) ⇒ val(sw.so, e)))

        healing: (G(∃sw. (cAct(sw) ∧ conn(sw.so,c.ci)) ∧ conn(p.po,sw.si)))

    }
    //Container for Architectural Guarantees
    ArchGuarantee {
        //Component Variables
        flex c : Car,
        flex p : Power

        //Data Type Variables
        flex e : Energy.energy

        //Configuration Trace Assertions
        delivery: (G(cAct(p)) ∧ val(p.po, e) ∧ cAct(c) ⇒ val(c.ci, e))
    }
}
```

*Figure 3.2: Textual Specification of the Scenario.*

As it could be observed in the textual specification of the model, in Figure 3.2, the architectural constraints detailing the behavior of the components with self-healing property are specified in the textual UI. First, a data type Energy is specified with only one sort declaration, similarly named 'energy'. Next, we see all the specification details of the three component types, Car, Power and Switch. Then, the important part of the behavior or configuration trace assertion specifications in terms of architectural constraints and guarantees are specified.

*Figure 3.3: Isabelle/HOL theory of the Scenario.*

Finally, in Figure 3.3, the assurance asserted by the architectural guarantee is transformed to Isabelle/HOL theory by the code generation engine. From the theorem a proof is sketched interactively and verification of the proof completed in Isabelle/HOL.

# Bibliography

(Rodosek, 2009) Rodosek, G. D., Geihs, K., Schmeck, H., & Stiller, B. (2009). Self-Healing Systems: Foundations and Challenges. Self-Healing and Self-Adaptive Systems, (09201).

(Broy, 2014) Broy, M. (2014, April). A model of dynamic systems. In Joint European Conferences on Theory and Practice of Software (pp. 39-53). Springer, Berlin, Heidelberg.

(Ascher, 2018) Ascher, D., & Kondzialka, C. (2018). Towards model-driven CIM-based data exchange for DSOs. Energy Informatics, 1(1), 23.

(Ascher, 2017) D. Ascher and D. Bytschkow. (2017). Integrating distribution system operator system landscapes. In Computer Science – Research and Development, Springer Berlin Heidelberg.

(Taylor, 2010) Taylor, R. N., Medvidovic, N., & Dashofy, E. M. (2010). Software Architecture: Foundations, Theory, and Practice.

(Ghosh, 2007) Ghosh, D., Sharman, R., Rao, H. R., & Upadhyaya, S. (2007). Self-healing systems—survey and synthesis. Decision support systems, 42(4), 2164-2185.

(TypeFox, 2017) TypeFox and Obeo, (2017) Xtext/Sirius - Integration: The Main Use-Cases [White Paper]

(Bettini, 2016) Bettini, L. (2016). Implementing domain-specific languages with Xtext and Xtend. Packt Publishing Ltd.

(Gidey, 2018) Gidey, H.K., Marmsoler, D. (2018). FACTUM Studio. https://habtom.github.io/factum/ [Code Repository]

(Broy, 2000) Broy, M. (2000). Algebraic specification of reactive systems. Theoretical Computer Science, 239(1), 3-40.

(Marmsoler, 2018, April) Marmsoler, D. (2018, April). Hierarchical specification and verification of architectural design patterns. In International Conference on Fundamental Approaches to Software Engineering (pp. 149-168). Springer, Cham.

(Marmsoler, 2017) Marmsoler, D. (2017, September). On the semantics of temporal specifications of component-behavior for dynamic architectures. In 2017 International Symposium on Theoretical Aspects of Software Engineering (TASE) (pp. 1-6). IEEE.

(Ballarin, 2003) Ballarin, C. (2003, April). Locales and locale expressions in Isabelle/Isar. In International Workshop on Types for Proofs and Programs (pp. 34-50). Springer, Berlin, Heidelberg.

(Marmsoler, 2017, October) Marmsoler, D. (2017, October). Towards a calculus for dynamic architectures. In International Colloquium on Theoretical Aspects of Computing (pp. 79-99). Springer, Cham.